\documentclass[12pt]{article}
\usepackage[english]{babel}
\usepackage{a4wide}
\usepackage{latexsym}
\date{}
\begin{document}

\title{\bf {\large{Special Holonomy Manifolds in Physics}}}
\author{\normalsize {Luis J. Boya}\\  
\normalsize{Departamento de F\'{\i}sica Te\'{o}rica, Universidad de Zaragoza} \\
\normalsize{E-50009 Zaragoza. Spain} \\
\normalsize{luisjo@unizar.es} \footnote{ \ \ To
  Jos\'{e} Cari\~{n}ena in his sixtieth birthday.}}

\maketitle
\begin{abstract}

This is a pedagogical exposition of holonomy groups
intended for physicists. After some pertinent definitions, we focus on
special holonomy manifolds, two per division algebras, and comment
upon several cases of interest in physics, associated with
compactification from $F$-, $M$- and string theory, on manifolds
of $8$, $7$ and $6$ dimensions respectively.\\

\end{abstract}

\vfill \eject

\section {Connections and Holonomy groups}

Let ($E, \nabla $) be a vector bundle with a {\it connection } $\nabla$
 : $P(M, G)$
is the principal bundle, and $F$ supports a representation of $G$, so $E(M,F)$
is the associated bundle:

\begin{equation}
 \nabla: \ \ F \ o\rightarrow E \rightarrow M
\end{equation}

$\nabla $ allows {\it covariant differentiation} of sections, e.g. for the tangent
bundle $E=TM, \nabla_X Y = Z$ means: covariant derivation of vector field $Y$ along 
(the flow of) $X$, is the vector field $Z$. In the general case $\nabla_X \psi =
\psi'$, where $\psi$, $\psi'$ are sections. $\psi: M \rightarrow E$. \\

Connections allow also {\it parallel transport} along paths. E.g., a frame 
$\underline {e}$ at a point $\cal P$  $\in M $ becomes
 another $\underline{e'} = g \cdot \underline{e}$ also at $\cal P$, after a loop
(closed path) $ \gamma $ through $ \cal P$, where $g \in G$. Consider all the loops 
from $\cal P$ and write $Hol(\nabla ):=\{g \}$; it is a (sub)group of $G$, called the
{\it holonomy group} of the connection; it was invented by E. Cartan in 1925. For
arcwise connected spaces, which is the case of manifolds, the holonomy group does 
not depend (up to equivalence) on the starting point $\cal P$.\\

Let $Hol_{0} (\nabla)$ be the restriction to {\it contractible} loops. Clearly
there is an {\it onto} map of the fundamental group
 
\begin{equation}
  \pi_{1}(M) \rightarrow Hol(\nabla)/Hol_{0}(\nabla)
\end{equation}

The restricted holonomy group $Hol_{0}(\nabla)$ is naturally connected, whereas 
$Hol(\nabla)$ needs not to be. For {\it generic} vector bundles the holonomy
group is expected to be as large as the structure group $GL(F)$. Now two important
theorems exist; first define \\

The {\it curvature} of a (vector bundle) connection $\nabla $ is the operator
on sections; e.g. vector fields:

\begin{equation}
R(X, Y):= [\nabla_{X}, \nabla_{Y} ] - \nabla_{[X, Y]}
\end{equation}

The curvature is  a {\it local } property, the holonomy a {\it global} one. 
But both are related be {\it the Ambrose-Singer theorem (1953)}:\\

	"The Lie algebra of the holonomy group is generated by the curvature". \\

The other grand result is called the {\it reduction theorem}:\\

	" The structure group can be reduced to the holonomy group".\\

That is, the total space of the bundle can be restricted by the holonomy loops.\\

	If the curvature is zero, the connection is said {\it flat}; the {\it restricted }
holonomy group $Hol_{0}(\nabla )$ is then $\{e \}$. Parallellelizable spaces
 ( $=$ trivial tangent bundle; they include $S^7$ and Lie groups)
 admit flat connections; just define the connection transport as translations
in the (trivializable) tangent bundle. \\

We shall consider mainly connections in the tangent bundle of a manifold;
then there is another tensor, the {\it torsion}, defined as

\begin{equation}
 T(X, Y):= \nabla_{X}Y - \nabla_{Y}X - [X, Y]
\end{equation}

Of course, the same space might have several inequivalent connections
(e.g. $S^3$ has the riemannian Levi-Civita connection, torsionless but curved,
and the Lie-group connection ($S^{3} \approx {SU(2)} $), flat but torsionful!).\\

We shall consider mainly {\it riemannian manifolds ($ {\cal V}, g$)}; 
they enjoy the standard {\it Levi-Civita} connection $\nabla = \nabla^g$ in the
 tangent bundle, which is symmetric and isometric:

\begin{equation}
(symmetric:)Torsion(\nabla ) = 0 = \nabla \cdot {g} (: isometric)
\end{equation} 

Let \underline{Isom} (${\cal V}, g$) be the isometry group of the manifold:
$ \ast \in $ \underline{Isom}, means $ g*=g$. A {\it generic} riemanifold has
no isometries, but the {\it generic} holonomy is the structure group, 
$O(n=dim \cal {V})$ or $SO(n)$. Spaces with maximal isometries 
 have constant curvature; for example \underline{Isom} $(S^n)=O(n+1)$, with
constant curvature $ K > 0$.\\

In physics both groups, isometry and holonomy, are important; for example,
in the {\it Kaluza-Klein} (de)construction, the {\it gauge} groups in the mundane
space ${\cal {V}}_4 $ come from the {\it isometry} group $U(1)$ of the 
compactification space $S^{1}$: that is why electromagnetism unfies with
gravitation with a circle as fifth dimension, so in this case ${\cal {V}}_{5} =
S^{1} \times {\cal {V}}_{4}$: the original Kaluza construction, 1919.\\

However, in the {\it supersymmetric} situation, it is the \underline{holonomy}
 group of the compactifiation space which fixes the number of supersymmetries;
for example, Calabi-Yau 3-folds ($CY_{3}$, real dimension $6$) are favoured for the 
dim $10 \rightarrow 4$ compactation of the Heterotic Exceptional string,
because the holonomy group, $SU(3)$, allows just ${\cal N}=1$ Susy in our
mundane, $4D$ space, as we want.\\

Isometries measure, of course, {\it symmetries}, whereas holonomy measures
distance (obstruction) from {\it flatness}; no apparent relation exists,
except opposite genericity (as stated above).\\

\underline{Simple examples} in $D=2$\\

The sphere $S^2$ has isometry $O(3)$, the torus $T^2$ has $U(1)^{2}$;
other genus $ g > 1 $ surfaces have {\it no} isometries. In the nonorientable
cases, $RP^{2}$ and Klein bottle are the only ones with isometries.\\

As for holonomy, the $2$-Torus $T^2$ is the only $CY_{1}$ among surfaces, because
is a group manifold, hence there is a connection with Hol $=\{e \}$ and $SU(1) \equiv
\{e \}$. The other surfaces with genus $ \neq 1 $ have Hol $=U(1)=SO(2)$ (if orientable)
and $O(2)$ (if not).\\

\underline{Simple examples } in $D=4$ \\

The "round" sphere $S^4$ has $O(5)$ as isometry, and a connection with $SO(4)$
holonomy. The $4$-Torus $T^4$ is flat, with isometry $U(1)^{4}$. Intermediate is
the topologically unique $K3$ (complex) surface (see later), which is a
 Calabi-Yau$_{2}$ space, with $dim_{R} =4$, with $SU(2)$ holonomy but no isometries.
As for $CP^2$, it has $U(3)$ as isometry group, and $U(2)$ for holonomy; in fact,
$CP^{2}\approx SU(3)/U(2)$.\\

	As introductory material, the first book on modern differential geometry is still
the best \cite{KN 1}. \\

Besides the original invention by E. Cartan (who did it in order to construct
all symmetric spaces ca. 1925/26), and a short revival in the fifties (Berger,
Lichnerowicz), the study of holonomy languished until resuscitation in the
mid-eighties, in part by imposition of physics (as in so many other mathematical
questions!). Then Bryant, Salamon and mainly Dominic Joyce (see the book
\cite {J 1}) revitalized greatly the subject.\\

Finally, let us note that the holonomy groups come to the world with a particular
 action (representation) in the tangent space, so one should properly speak of 
the holonomy {\it representation}. \\

\section{Special and Exceptional Holonomy Manifolds}

What groups can appear as holonomy groups $Hol(g) \subset O(n)$ of 
riemanifolds ($\cal {V}$$_{n}, g$)? The issue was set and solved by
M. Berger in 1955. To state precisely the problem, suppose $Hol(\cdot )$
acts {\it irreducibly} in the tangent space, and {\it symmetric} spaces
$G/H$ are excluded (because all are known (Cartan) and for them the subgroup $H$
is the holonomy group). Berger found all possible candidate groups with
these prescriptions by a hard case-by-case method.\\

	Berger\'{}s solution is best understood (Simons, 1962) as
{\it the search for transitive group actions over spheres}:
with two exceptions, these are the special holonomy groups.\\

The generic case is the orthogonal group acting {\it trans} on the
 sphere, $O(n) \circ \rightarrow S^{n-1}$, with isotopy
$O(n-1)$, that is $S^{n-1} = O(n)/O(n-1)$. The cases of $trans$
action on spheres coincident with special holonomy manifolds are
(Berger\'{}s list):\\

\begin{equation}
\begin{array}{cccccc}
R &  O(n) & or & SO(n) & acting \ \ on & S^{n-1}  \\
C &   U(n) & or & SU(n) & acting \ \  on & S^{2n-1}  \\
H &   Sp(n) \cdot Sp(1) & or & Sp(n) & acting \ \ on & S^{4n-1} \\
O & Spin(7) & on \ \ S^{7} & or & G_{2} & on \ \ S^{6} 
\end{array}
\end{equation} \\

We exhibit the association with the four division
algebras $R$, $C$, $H$, y $O$, which is obvious and remarkable.
Recall also that the homology of compact simple Lie
groups is given by that of the product of odd-dimensional
spheres, see e.g. \cite{B 1}. Then the real and complex cases
are clear, for example $SU(3) \approx S^{3} \times S^{5}$, as homology
sphere product, so we have $S^5 = SU(3)/SU(2)$. $Sp(n)$ for us
is the compact form of the $C_{n}$ Cartan Lie algebra. Also
there is a "nonunimodular" form

\begin{equation}
Sp(n) \cdot Sp(1) := Sp(n) \times_{/2} Sp(1)
\end{equation}

As for the octonion cases, recall dim $Spin(7) = 8$,
 type (+1, real); in some sense which we do not elaborate,
it could be said that $Spin(7)$ "is" $Oct(1)$, and $G_{2}$, defined
as $Aut(O)$, is the "unimodular" form, $G_{2} \approx SOct(1)$.\\

There are two more cases of {\it trans} actions on spheres

\begin{equation}
Sp(n) \cdot U(1) := Sp(n) \times_{/2} U(1) \ \ acting \ \ in \ \ S^{4n-1}
\end{equation}
and

\begin{equation}
Spin(9) \ \ acting \ \ in \ \ S^{15}
\end{equation}

which, however, do {\it not} give rise to new holonomy groups.
$Spin(9)$ acts {\it trans } in $S^{15}$ as $Spin(9) \approx S^3 \times S^7
\times S^{11} \times S^{15}$. In fact, $S^{15} = Spin(9)/Spin(7)$, equivalent,
in some sense, to $S^{15} \approx "Oct(2)"/"Oct(1)"$. Spin(9) was really
in Berger\'{}s list, but the only space found was $OP^2$ (Moufang or
octonionic plane), which is a {\it symmetric} space.\\

The sphere $S^7$ of unit octonions is singularized
 because there are {\it four} groups with {\it trans} actions, $O(8),
 U(4), Sp(2)$ and $ Spin(7) = "Oct(1)"$; similar for $S^{15}$, but no
more.\\

	Notice the next Spin case, $Spin(10)$: the action is {\it not
 trans} in the higher sphere, to wit, $dim \ \ Spin(10) = 16$, complex, so
$Spin(10)$ acts on $S^{31}$, but the sphere
 homology product expansion for $O(10)$ is $S^{3}\times S^{7}\times S^{11}
\times S^{15}$.\\

We expand now on the extant cases: 	\\

\underline{Over the reals} we have the groups $O(n)$, generic
holonomy, and $SO(n)$: clearly the second obtains
 when the space is orientable and the connection oriented:
there is an obstructoin, the first Stiefel-Whitney class:

\begin{equation}
{\cal V} \ \ \  orientable \ \ manifold \\ \leftrightarrow w_{1}=0, w_{1} \in H^{1}({
\cal V},Z_{2})
\end{equation}

Alternatively, the manifold $\cal V$ should have a global
volume element (reduction $GL(n, R) \leftarrow SL(n, R))$. In
Berger\'{}s classification, he took the manifolds as simply
connected, which are then automatically orientable ( if
$\pi_{1}({\cal V})=0$, all first order (co)homology vanishes,
including $w_{1}$). Hence, $O(n)$ did not appear in his list.\\

\underline{Over the complex} we have complex manifolds, with 
structure group $U(n)$; but a generic hermitian metric $h=
g+i\omega$ will allow in general a connection with holonomy $SO(2n)$, as
$\nabla g=0$ only, unless the complex structure $J$ is also preserved:
 this is the case of 
K\"{a}hler manifolds, with $\nabla \omega (= d\omega)=0$,  where 
 $\omega =g(J)$
is the symplectic form.\\

	The "unimodular" restriction $SU(n)$ obtains when
the associated bundle with group $U(1)=U(n)/SU(n)$ is
trivial, which is measured by the first Chern class:

\begin{equation}
SU(n) \ \ holonomy \ \ \leftrightarrow 0=c_{1} \in H^{2}({\cal V}, Z)
\end{equation}

The natural name for these spaces would be "Special
 K\"{a}hler manifolds", but had become known instead as
"Calabi-Yau spaces", after the conjecture of E. Calabi
proven by S.T. Yau. As a bonus, these spaces have trivial
Ricci tensor ($Ric = Tr \ Riem$, contraction of the Riemann tensor):
define the Ricci $2$-form $\rho := Ric(J)$; by the same token as
above $J$ parallel (=covariant constant) implies $\rho$ closed,
and it turns out that $[\rho ] = 2\pi c_{1}({\cal V})$. Therefore

\begin{equation}
SU(n) \ \ holonomy \ \ \ implies \ \ Ricci \ \ flat \ \ manifolds, \ \ Ric=0.
\end{equation}

In other words, Calabi-Yau spaces are candidates to Einstein spaces, solution
of vacuum Einstein equations (without cosmological term). Also $CY$ spaces
have an holomorphic volume element.\\

\underline{Over the quaternions} we have quaternionic
manifolds, like $HP^n$, with "nonunimodular" holonomy
$Sp(1)\cdot Sp(n)$, and hyperk\"{a}hler manifolds, with
holonomy $Sp(n)$. As clearly $Sp(n)\subset SU(2n)$, the later are
also Ricci flat. In fact, the space $HP^{n}=Sp(n+1)/Sp(1)\cdot Sp(n)$,
which a quaternionic space, is not valid
 as special holonomy manifold, because it is a
symmetric space. By contrast, quaternionic manifolds
have not to be even K\"{a}hler, as the example of $S^4 = HP^1$
shows; quaternionic and hyperk\"{a}hler manifolds are not easy to come by. \\

\underline {Manifolds with {\it octonionic} holonomy}  
: The two groups: $Spin(7) \subset SO(8)$, acting as holonomy groups
on 8-dim manifolds, and $G_{2} \subset SO(7)$, in 7-dim. manifolds, are
groups associated to the octonions. $Spin(7) \approx "Oct(1)"$ and
$G_2 = \ Aut(O) \approx "SOct(1)"$; manifolds with these holonomy groups
are called {\it exceptional holonomy} manifolds. 
We expect them to play some role in physcs, as for example the internal
spaces in $F-$ and $M-$theories have dimensions $8$ and $7$ bzw.\\

	It is remarkable that, while $O$, $U$ and $Sp$ are isometry
groups of symmetric regular positive bilinear forms ($O$),
sesquilinear regular positive forms ($U$) and antisymmetric
regular forms ($Sp$), the octonionic cases obtain from
invariance of certain $3-$forms ($G_2$) and selfdual $4$-forms
($Spin(7)$). This cannot go beyond dimension $8$, because
for example the dimension of $3$-forms in $9$-dimensions is
${9 \choose 3} =84$, whereas dim $GL(9, R)=81$.\\

	In fact, the algebraic definition of $Spin(7)$ is the
isotopy group of {\it certain} class of self-dual $4$-forms
in $R^4$: it preserves also orientation and euclidean metric, so
$Spin(7) \subset SO(8)$. Notice the selfdual form is {\it not}
generic, as $8^{2} - {8 \choose 4}/2 = 29 > dim \ Spin(7) = 21$;
the special selfdual four-form is called the {\it Cayley form}
in the math literature. In any case $Spin(7)$ covers $S^7$ with
isotopy $G_{2}: \ 21 - 7 = 14$.\\

The {\it algebraic} definition of $G_2$ is this: the stability
group of the {\it generic} $3$-form in $R^7$ as vector space:
dim $GL(7, R) -  dim \wedge ^{3} R^{7} = \ \ dim \  G_{2}: 49 -35 = 14$.
Of course, the original characterization of $G_2$ as $Aut(O)$ by
Cartan is related to this: a $3$-form becomes a $T^{1}_{2}$ tensor through
a metric, and this is indicative of an algebra, i.e. a bilinear map
$R^{7} \times R^{7} \rightarrow R^7$, given by the octonionic product
(and restriction to the imaginary part).Indeed, alternativity of the
octonion product corresponds to antisymmetry in the 3-form.\\

There is also a sense in which for each division algebra there is a
{\it normal} form and am {\it unimodular} form:\\

\begin{equation}
\begin{array}{ccc}
Reals & O(n) \approx S^{0} \times S^{3} \times S^{7}...
 & SO(n) \approx S^{3} \times S^{7}...\\
R & Generic & Orientable, \ \ w_1=0 \\
 & & \\
Complex & U(n) \approx S^{1} \times S^{3} \times S^{5}... 
& SU(n) \approx S^{3} \times S^{5}...\\
C & K\ddot{a}hler & Calabi - Yau, \ \ c_{1}=0\\
 &  & \\
 Quaternions & Sp(n) \cdot Sp(1) \approx S^{3} \times S^{3} \times S^{7}...
& Sp(n) \approx S^{3} \times S^{7}... \\
H & Quaternionic & Hyperk\ddot{a}hler \\
  &  & \\
Octonions & Spin(7) \approx S^{3} \times S^{7} \times S^{11} & G_{2} \approx S^{3}
 \times S^{11} \\
O & dim \ \ 8 & dim \ \ 7   
\end{array}
\end{equation}

\underline{NOTES} 1). Today there are {\it compact} examples of all
cases of special holonomy manifolds: big advances were made recently
by Joyce \cite{J 1}, Salamon and others.\\

	2). In dim $4$, a remarkable case is the $K3$ manifold or Kummer (complex)
surface (the name is due to A. Weil, 1953, for Kummer, K\"{a}hler and
Kodaira). It is the only $CY$ manifold in dimension four; it can be easily 
constructed ($R^{4} \rightarrow T^{4} \rightarrow \ \ Z_2 \ \ orbifold \rightarrow \ \
blow \ \ up$; \cite{CG} ). For a long time it was the paradigmatic example of
$SU(n=2)$ special holonomy.\\

3). Notice a generic complex $n$-manifold would have $SO(2n)$ holonomy
inspite the structure group being $U(n)$!\\

4). The Calabi conjecture, proved by Yau, indicates the relation of the
Ricci form with the K\"{a}hler structure.\\

5). In the 80s a big industry, led by Phil Candelas in Austin ( \cite{PhC}), was to find
$CY_{3}$ manifolds for string compactifications. {\it Mirror} symmetry
was discovered in this context; see later. \\

6). Except $G_{2}$, all special holonomy spaces are even dimensional.\\

7). One can show that $G_2$ and $Spin(7)$ holonomies are Ricci flat.\\

8). Although special holonomy representations are irreducible in the vector case, 
there might be $p$-forms which split under the holonomy subgroup. For example, for
$G_{2}$, $3$-forms split as $35 = 1 + 7 + 27$; the $7$ irrep is justly the octonion
product, and the $1$ the invariant $3$-form. As for $Spin(7)$, a self-$4$-form
splits as $35 = 1+7+27$: it includes the invariant 4-form. 

\section{Cases in Physics: dimensions $6$, $7$ and $8$}.

	In 1983, just after the first studies in eleven dimensional
supergravity (11-dim SuGra), Duff and Pope realized that it is
the {\it holonomy} of the compactified space which determines
the number of surviving Susy symmetries down to $4$ dimensions. 
For spinor fields, as $S^{7}= Spin(7)/G_{2}$, $7$-manifolds 
with exceptional $G_{2}$ holonomy would have a surviving spinor,
hence ${\cal N} = 1$ Susy down to $4$-D. But after the String
Revolution, 1984/85, the descent $10 \rightarrow 4$ took over,
 and the favourite spaces were $CY \ 3$-folds: the heterotic string
has ${\cal N} = 1$ supersymmetry in $10-$dim., which means ${\cal N} = 4$ 
down to earth; but it will be $1/4$ of these after $CY_{3}$ compactation:
the generic $SO(6)$ holonomy of any (orientable) dim-$6$ manifold would
become $SU(4)=Spin(6)$ after imposing a (necessary) spin structure, and
then if we want one spinor to survive the group descends to $SU(3)$. The
$6$-dim manifold has to be orientable, spin, complex, K\"{a}hler and
Calabi-Yau. See \cite{ Duff}  \\

With the advent of $M$-Theory (1995), P. Townsend resurrected the
idea of $7$-dim manifolds with $G_{2}$ holonomy. One can go even further to
Spin($7$), the largest exceptional holonomy group, by considering for example
compactifications to $3$-dim spaces (which seems natural; for example, the series
of noncompact symmetries of supergravity includes $E_{7}$ in $4$ dimensions, which
is claiming for $E_{8}$ in three, which is of course the case). Another reason is
$F$-Theory, which works in $12$ dimensions with ($2, 10$) signature, and where
Spin($7$) (perhaps in a nonpositive form) fits well. \\

	Are such beasts as $CY_{3}$ spaces in abundance? Yes, you can produce them in
assembly line, to the point of studying their Hodge numbers statistically! 
\cite {PhC}. Another interesting phenomenological constraint in the ``old-fashion''
$10 \rightarrow 4$ descent, was the Euler number $\chi $ : it is related, via zeroes
of the Dirac operator, to the number of generations, 
which is $\mid \chi \mid$/$2$.\\ 

	As for the extension to $F$-Theory, we refer the reader to \cite{LJB}. 
Besides some attempt to state the particle content, the theory is rather
stagnant at this point (as is $M$-Theory in general). 
For a modern study of special holonomies with Lorentzian metrics, see \cite{Bry}.\\

\section{Relation with Mirror Symmetry}

	Complex manifolds have a refinement on Betti numbers, as they separate
in holo- or antiholomorphic. The full expression of them is called the
\underline {the Hodge diamond}. For example, for the previous $K3$ 
manifold it is

\begin{equation}
\begin{array}{ccccc}
 & & h^{0,0}=1 & &  \\
 & h^{1,0}=0 & & h^{0,1}=0 & \\
h^{2,0}=1 & & h^{1,1}=20 & & h^{0,2}=1 \\
  &   h^{2,1} = 0 & & h^{1, 2} = 0 & \\
  &  & h^{2,2} = 1 &  &  
\end{array}
\end{equation} \\
with bettis $= 1, 0, 22, 0, 1 $. For Calabi-Yau $3$-folds the diamond is bigger, but still symmetric. A {\it mirror}
pair $X, Y$ of $CY_{3}$ are two such spaces with

\begin{equation}
h^{1, 1}[X] \ \ and \ h^{2, 1}[X] \ \ \ equal \ \ h^{2, 1}[Y] \ \ and \ h^{1, 1}[Y]
\end{equation} 

There is no clear reason for this duality, but just another more example of
a physics discovery on pure mathematics. These two numbers measure very different
invariants, so mirror symmetry came up as a big surprise to mathematicians, when
many conjectures by physicists seemed to be true. \\

Indeed, several of these conjectures {\it were} true. It is also true
that string theory "glosses over" orbifold singularities (i.e. quotienting
manifolds by fix-point-action discrete groups), and the associated (quantum)
conformal field theories make perfect sense. \\

We shall only comment on the interpretation of this
mirror symmetry by M. Kontsevich \cite{Kon}. For him, the crux of the matter
is a trade between {\it complex} and {\it symplectic} geometry. In fact, complex
structures preserve an imaginary unit, $J \in End \ {\cal V}, \ J^{2} = -1$, whereas
a symplectic manifold enjoys a $2$-form $ \omega$, whose matrix form is very 
similar to $J$!. In both cases, there is an extra condition: \\

In the complex case, $N(J)=0$, where $N$ stands for the Nihenhuis (obstruction); 
that makes up a complex, not only almost-complex, manifold
 (the Newlander-Nirenberg theorem). In the symplectic case, the $2$-form is closed,
or, alternatively, the (inverse) Poisson bracket satisfies Jacobi\' {}s identity.
Their isotropy groups, $GL(n, C)$ and $Sp(n, R)$ respectively, are of the same
homotopy type, namely the homotopy of the intersection, $U(n)$. In any case,
the relation hidden in Mirror Symmetry is an intrincate one. In the words of
Dijkgraaf: "Mirror symmetry is the claim that the generating function for
certain invariants of the symplectic structures on the
 $2$-Torus $S^{1} \times S^{1}$ is a "nice" function in the moduli space of complex
structures in the same": the $2$-Torus is a {\it self-dual} manifold for Mirror
Symmetry (MS).  \\

	The general case of MS is best understood in terms of toric varieties, which
generalize projective spaces.\\

	The main lesson of MS for physics seems to be this: certain topology changes
(sometimes called "flops") are compatible with the underlined string theory.
That probably means that the complex-geometric description of strings is too
fine... Perhaps it hints towards a new type of duality. \\

\vfill \eject

\end{document}